\documentclass[sigconf, nonacm, natbibopts={numbers}]{acmart}

\makeatletter
\renewcommand{\acmBadgeL}[2][]{}
\renewcommand{\acmBadgeR}[2][]{}

\usepackage{float}
\usepackage{amsmath}
 
\usepackage[english]{babel}
\usepackage{blindtext}
\usepackage{listings}

\AtBeginDocument{%
  }

\setcopyright{none}             
\makeatletter
\let\@acmprice\@empty            
\let\@doi\@empty                 
\renewcommand{\@formatdoi}[1]{}  
\let\@journalCodeAndYear\@empty
\let\@bibstrip\@empty
\let\printcopyrightnotices\@empty
\renewcommand{\@copyrightpermission}{}

\def\@journalandvolume{}
\makeatother

\setcopyright{none} 
\renewcommand\footnotetextcopyrightpermission[1]{} 
\begin{document}

\title{A scalable online machine learning approach for Stock Recommendation}

\author{Harsh Nagarkar}

\email{harshnagarkar27@gmail.com}

\affiliation{%
  \institution{California State University Chico}
  \streetaddress{400 W 1st St.}
  \city{Chico}
  \state{California}
  \country{USA}
  \postcode{95928}
}



\maketitle

\section{Abstract}
Stock recommendation systems face the dual challenge of adapting to rapidly changing market conditions while maintaining low-latency predictions for end users. Traditional batch-trained models fail to capture concept drift, and monolithic architectures struggle to provide fault tolerance under load. This paper presents a scalable, online deep learning-based stock recommendation system built on a distributed microservices architecture using Kubernetes, Docker, and RabbitMQ. The system employs a hybrid leader-follower architecture where a primary model continuously trains on streaming financial data — including EPS, MACD, and price from the Alpha Vantage API — while multiple replica models serve user-facing recommendations in parallel. A multi-layer perceptron implemented with TensorFlow Recommenders generates content-based recommendations using explicit user ratings (1–5) and transfer learning. The architecture ensures high availability: the leader persists model weights to Google Cloud Object Storage, allowing replicas to recover seamlessly upon failure, while RabbitMQ provides message durability and replay. Results demonstrate that the system serves stock recommendations in 2–3 seconds per request and processes up to 500 portfolio addition requests per second per follower. Key limitations include data staleness (up to 150 minutes due to API rate limits) and the absence of a service mesh for inter-cluster security. This work contributes a production-ready reference architecture for online recommender systems that balances consistency, availability, and scalability in a financial domain context.

\section{Contributions}
This work makes several contributions to the design and implementation of online recommendation systems for financial portfolios.

First, we propose a novel hybrid leader-follower architecture for online recommender systems that decouples continuous training from low-latency serving. In this design, a single leader model trains incrementally on streaming data and persists updated weights to cloud object storage after each observation, while stateless follower replicas serve user-facing recommendations in parallel without disk writes. This separation enables horizontal scaling of inference under load without compromising model freshness, and provides a clear failure recovery path: followers reload the latest leader weights upon restart, and RabbitMQ message durability ensures no data loss during downtime.

Second, we demonstrate that an online deep learning model using explicit user ratings (1–5) can effectively personalize stock recommendations and capture concept drift in user preferences over time. Unlike existing financial recommender systems that rely on implicit signals or batch training, our approach updates predictions continuously as users rate stocks, allowing the model to adapt to changing risk tolerance and market conditions without full retraining.

Third, we show that meaningful stock recommendations can be generated using only five parameters — EPS, MACD-K, MACD-S, current price, and stock name — suggesting that near real-time portfolio management is achievable without extensive feature engineering. This lowers the barrier for personalized investment tools and indicates that online learning models can extract sufficient signal from a minimal feature set.

Fourth, we provide a production-ready reference architecture built entirely on open-source, cloud-native technologies (Kubernetes, Docker, RabbitMQ, PostgreSQL,
TensorFlow Recommenders) with empirically measured performance characteristics: 2–3 second recommendation latency, 300 portfolio addition requests per second per follower, and documented consistency-availability tradeoffs. Each identified bottleneck is mapped to a concrete remediation path, offering practitioners a grounded upgrade roadmap.

Finally, we document the practical limitations and failure modes encountered during implementation — including leader recovery semantics, cold start behavior, and the absence of service mesh security — providing transparency about the system's operational boundaries and guiding future research directions.

\section{Introduction}
This paper presents the implementation of an online deep learning recommender system for stock recommendations. The system generates personalized stock suggestions using a deep learning model built on TensorFlow's Recommenders library \cite{tensorflow2015-whitepaper}, incorporating explicit user feedback through a content-based recommendation approach.

This work is inspired by two key projects. The first is Uber's "Michelangelo" framework, a machine learning platform developed for real-time ride predictions \cite{hermann_2017}. The second is "A Personalized Stock Recommendation System using Adaptive User Modeling," which employed implicit user feedback and was implemented in Java \cite{4141428}. In contrast, our implementation is written in Python and deployed on a distributed cluster architecture using Kubernetes \cite{10.5555/3175917}, Docker \cite{merkel2014docker}, and RabbitMQ \cite{10.5555/3126138}. Kubernetes, provided by Google Cloud Platform, serves as the microservice management framework; Docker provides application containerization \cite{merkel2014docker}; and RabbitMQ \cite{10.5555/3126138} handles message queuing.

Several challenges arose during development. Ensuring data consistency across multiple machine learning follower replicas and the leader proved non-trivial, as did managing the queuing of incoming data for processing and presenting the results to the user in an understandable format. Despite these hurdles, this paper presents a fault-tolerant system for stock recommendation that addresses these concerns. We also discuss the inherent trade-offs in this approach, particularly bottlenecks related to dataset update frequency and maintaining acceptable user interaction latency.

\section{Background and Related Work}

Recommendation systems are broadly classified into two main categories: content-based recommendation and collaborative-based recommendation. Collaborative filtering requires a large number of users interacting with the system to generate meaningful recommendations. Since this project is limited in the number of users, we employ content-based recommendation. Additionally, stock risk tolerance varies significantly between individuals, making content-based approaches more suitable by tailoring recommendations to each user's specific data features \cite{ZHANG2018223}. In content-based recommendation, an item is recommended to a user based on the feature attributes available in the dataset.

The List-wise Recommendation (LIRD) model proposed in "Deep Reinforcement Learning for List-wise Recommendations" \cite{zhao2017deep} recommends items through online interaction, updating recommendations as the user interacts with the model within a given time window. However, LIRD relies on a Markovian decision-making process. In contrast, our implementation uses a simple deep neural network — a multi-layer perceptron — to model the non-linear relationships in stock rating data. Our model is equally capable of capturing concept drift, which LIRD also attempts to model. A key difference is that our system implements an explicit rating-based mechanism where users vote from 1 to 5 on stocks, whereas LIRD relies on click counts and other implicit signals alongside some explicit factors. The entire model updates recommendations continuously
as users interact with the system and the dataset is refreshed. We have designed an online recommendation system that is scalable and does not remove pre-existing stocks from recommendations, since the dataset is updated frequently. Currently, the update interval is set to 150 minutes for prototyping, though ideally it should be a few seconds. Our problem statement is: "How to build an online stock recommendation system that is also scalable."

The remainder of this paper is organized as follows. Section 3 discusses the system architecture and the model's inner workings. Section 4 describes the user interface and how recommendations are presented. Section 5 outlines the limitations of the current approach. Section 6 discusses future work, and Section 7 concludes the paper.

\subsection{Stock Data Analysis and Inputs}

Stock analysis for recommendation can be approached in several ways, but it is fundamentally divided into three categories: fundamental data analysis, technical data analysis, and combined data analysis \cite{nti_adekoya_weyori_2019}. Many organizations use SWOT (Strengths, Weaknesses, Opportunities, and Threats) analysis as a combined qualitative and quantitative technique. However, for the purposes of this paper, we focus on quantitative fundamental data analysis \cite{nti_adekoya_weyori_2019}. This information is obtained from 10-K and 10-Q forms submitted to the SEC, which contain data such as cash flow, revenue, and earnings per share (EPS). Fundamental analysis is typically employed by long-term investors seeking to capture underlying company value \cite{nti_adekoya_weyori_2019}. In contrast, technical analysis is used by short-term traders who rely on indicators to capture market trends \cite{nti_adekoya_weyori_2019}. These indicators — including moving average convergence divergence (MACD), relative strength index (RSI), and volume-weighted average price (VWAP) — provide insight into buy and sell signals. See Figure \ref{fig: Data fed into
the model}.

For this implementation, we use the most commonly traded indicators: MACD and EPS. In future work, we plan to incorporate RSI, revenue, and cash flow to improve prediction accuracy.

\begin{center}
\begin{figure}
\includegraphics[scale=0.35]{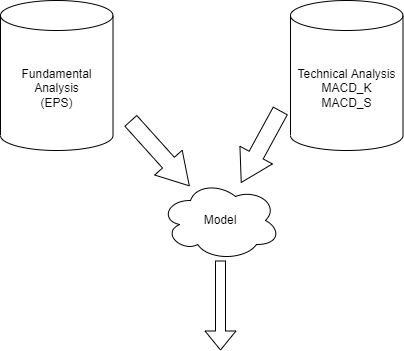}
\Description{Data that is passed into the model}
\caption{Data that is passed into the model}
\label{fig: Data fed into the model}
\end{figure}
\end{center}

\begin{itemize}

\item {\texttt{EPS:}} Earnings Per Share (EPS) represents the portion of a company's profit allocated to each outstanding share of common stock. It serves as a profitability indicator: higher profits yield higher EPS, making it a key performance metric for investors.

\bigskip
\begin{center}
\[ X = \frac{\text{Net income after Tax}}{\text{Total number of shares}} \]
\cite{nti_adekoya_weyori_2019}
\end{center}

\item {\texttt{MACD:}} The Moving Average Convergence Divergence (MACD) is a technical indicator calculated from the relationship between two functions. The first, MACD-K, is the difference between the 26-day and 12-day exponential moving averages. The second, MACD-S, is the 9-day exponential moving average calculated using the same formula as MACD-K but without the final term. The intersection or convergence of these two lines indicates buy or sell signals, capturing the trend of a stock in the market.

\bigskip
\begin{center}
\[ \mathrm{MACD} = \sum_{k=1}^{N} \mathrm{EMA}_k - \sum_{d=1}^{N} \mathrm{EMA}_d \]
\cite{nti_adekoya_weyori_2019}
\end{center}
\bigskip

\end{itemize}
These performance indicators, along with price and stock ticker, are fed into the machine learning model for training.

\subsection{Data Collection}

The dataset for this model is limited to S\&P 500 stocks, though the approach is theoretically scalable to the entire stock market. Data is continuously gathered using the Alpha Vantage Pro API \cite{vantage}, which provides precomputed values for EPS, MACD-K, and MACD-S, along with the latest 15-minute closing price for each stock ticker. Due to budget constraints, we use the Alpha Vantage Pro API \cite{vantage} service plan, which allows 75 API requests per minute. This limits stock updates to every 90 minutes. With a higher-tier plan, the database could be updated every 3 to 5 seconds, which is identified as future work. The highest consumer plan supports 1,200 requests per minute, and enterprise-scale plans are available upon request. Additionally, the Alpha Vantage Pro API \cite{vantage} does not cover all S\&P 500 stocks; we successfully collected data for 498 of the 500 stocks \cite{datopian}.

\section{Architecture}

The system architecture was designed from a high availability standpoint. The primary challenge lies in reliability — specifically, the synchronization of machine learning model instance clones across multiple Kubernetes pods. The ML model follows a leader-follower pattern. From an online machine learning perspective, followers use incoming observations to train themselves without storing those observations. The leader \cite{fern_givan} functions similarly.

\begin{figure}
 \centering
 \includegraphics[scale=0.35]{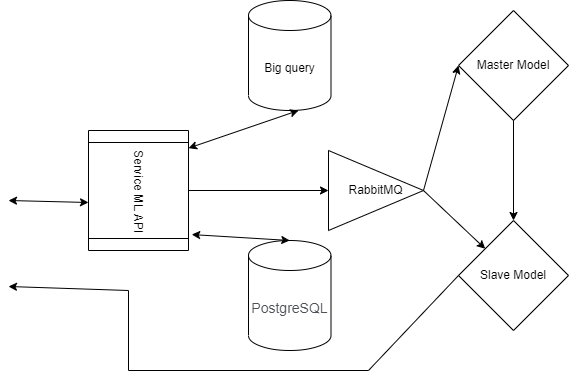}
 \caption{Architectural Design}
 \label{fig:Data flow and Architecutral design}
\end{figure}

While leader failure is rare — the Kubernetes cluster priority is set to the maximum preemption limit — it remains a possibility. To ensure users can always view their portfolios, hand-picked stocks are stored as data logs in PostgreSQL. In the event of leader failure, the model can be retrained to restore weights, though the recovered state will not be identical to the pre-failure state. There would also be data discrepancies if followers scale during this recovery window. A voting mechanism where a follower could automatically assume the leader's role upon failure has not yet been implemented and is left as future work. See Figure \ref{fig:Data flow and Architecutral design}.

\subsection{Data Ingress and Egress}

The system has three data endpoints. The first is the Alpha Vantage Pro API \cite{vantage}, which fetches stock data via HTTP requests and stores it in the database. The second endpoint connects to the frontend: upon receiving GET or POST requests, it either returns a database query result or places data into RabbitMQ for further processing. The third endpoint serves recommendations to the user. Together, these three endpoints handle all ingress and egress communication for the system.

\subsection{Data Processing and Transfer}

Data passed to RabbitMQ fans out to both the leader and follower models simultaneously for processing. Data that does not belong to the machine learning model is handled via direct database queries and returned to the user. When new followers need to be spawned for replication, they load the latest weights from Google's Object Storage \ref{fig:Data syncing between leader and follower}.

\subsection{Model Implementation}

A deep learning approach was chosen for recommendations because a multi-layer neural network can effectively model non-linear data. The recommendation model operates in a hybrid architecture combining two processing strategies: (1) leader-follower pattern processing, where the leader writes weights and followers read them, and (2) parallel processing, where the same data passes through both leader and follower simultaneously. Both components are attached to Google's Object Storage via a FUSE adapter for synchronization. The following subsections describe the role of each component.

\subsubsection{Model Architecture}

\begin{itemize}

\item \texttt{\textbf{Leader:}} There is a single leader that processes each data query. Upon completion, it updates the model weights and writes them immediately to Google's Object Storage. In the event of leader failure, the leader queries past observations from the PostgreSQL log and retrains the model, subsequently rewriting the weights. Leader
failure is mitigated by setting its Kubernetes preemption priority to the maximum. RabbitMQ also ensures that data is queued during downtime and replayed from the point of failure when the leader recovers.

\begin{figure}
 \centering
 \includegraphics[scale=0.35]{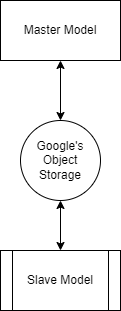}
 \caption{Data syncing between leader and follower}
 \label{fig:Data syncing between leader and follower}
\end{figure}

\item \texttt{\textbf{Follower:}} Multiple followers process the same data queries as the leader, but they do not write data to disk. This makes them faster and keeps them updated throughout the model's runtime. If a follower fails, a new instance reloads the latest computed weights from the leader's saved model location in Google's Object Storage, provided the leader is not behind. By pre-scaling followers to meet expected demand, random follower failures can be avoided. A caveat exists if the leader falls behind — in this scenario, a follower may miss some observations upon restarting after failure, though this is considered an edge case. From an availability perspective, the system remains operational throughout this process \ref{fig:Data syncing between leader and follower}.

\end{itemize}

\subsubsection{Generating Recommendations}

\begin{itemize}
\item \texttt{\textbf{Model Development:}} The model is built on the TensorFlow Recommenders library and adapted to the project's requirements. Text data is processed using the \texttt{StringLookup} function, which converts strings to integer indices; an embedding layer then converts these into dense tensors. Numerical values are processed in two ways: (1) they are discretized into binary representations in Hamming space \cite{le2021efficient} and then embedded, and (2) they are normalized for direct use. The recommendation system operates via transfer learning, training on a specific set of tensors to make predictions on unsupervised domain data \cite{pan2009survey}.

\item \texttt{\textbf{Candidate Generation:}} This is a multi-tasking model. It first rates unsupervised data using an artificial neural network, then compares these ratings against known ratings during training. The rating model is defined as follows:

\begin{lstlisting}[breaklines]{python}
rating_model = tf.keras.Sequential([
    tf.keras.layers.Dense(256, input_shape=(328,), name="dense1"),
    tf.keras.layers.Dense(128, activation="relu", name="dense2"),
    tf.keras.layers.Dense(1, name="dense3"),
])
\end{lstlisting}

The ranking task uses mean squared error loss and root mean squared error as the evaluation metric:

\begin{lstlisting}[breaklines]{python}
tfrs.tasks.Ranking(
    loss=tf.keras.losses.MeanSquaredError(),
metrics=[tf.keras.metrics.RootMeanSquaredError()],
)
\end{lstlisting}

A retrieval layer generates feature representations for stocks, using factorized top-K candidates:

\begin{lstlisting}[language=Python]
tfrs.tasks.Retrieval(
    metrics=tfrs.metrics.FactorizedTopK(
        candidates=detailed_stock.batch(1)
    )
)
\end{lstlisting}

To retrieve the top 25 recommended stocks, a brute-force search is performed over the candidate embeddings:

\begin{lstlisting}[breaklines]{python}
brute_force = tfrs.layers.factorized_top_k.BruteForce(
    model.query_model
) 
brute_force.index_from_dataset(
    detailed_stock.batch(128).map(
        lambda title: (title, model.candidate_model(title))
    )
)

data, stocks_recommended = self.brute_force(
    np.array([username]), k=25
)
\end{lstlisting}

\begin{figure}
 \centering
 \includegraphics[scale=0.4]{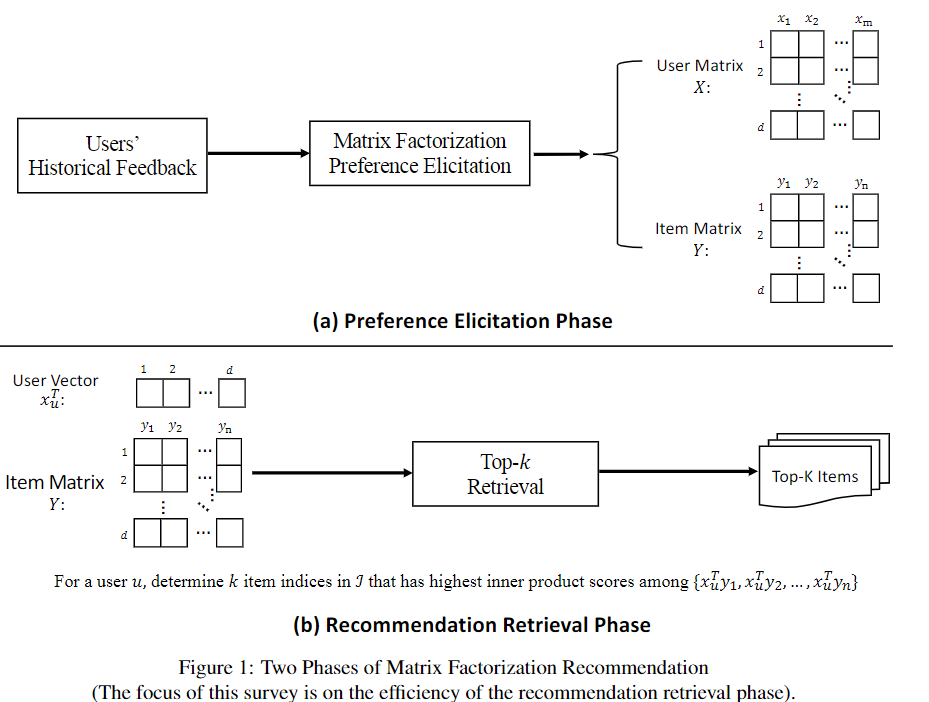}
\caption{Retrieval candidates based on query \cite{le2021efficient}}
 \label{fig:Candidate Retrival}
\end{figure}

\item \texttt{\textbf{Epochs:}} The model trains for three epochs on each new data observation.

\item \texttt{\textbf{Optimizer:}} The optimizer is Adagrad with an initial learning rate of 0.5.

\item \texttt{\textbf{Rating Process:}} Users can rate stocks from 1 (least liked) to 5 (most liked).

\item \texttt{\textbf{Concept Drift:}} The online learning model continuously adapts to new observations and gradually forgets older ones. This allows the system to capture evolving user interests and recommend stocks based on the most recently liked items.

\end{itemize}

\subsection{Data Storage}

The large volume of stock data, updated at a relatively slow rate, is stored in a BigQuery data lake. BigQuery provides cost-effective storage for large datasets while supporting SQL queries. A key feature is CSV batch upload, which allows massive data insertion in a single job. User portfolio data, on the other hand, is stored in PostgreSQL, where it is updated each time a user adds a stock to their portfolio.

\section{User Interface}

The user interface presents all information in a consumable format. The frontend is built with React and Tailwind CSS, providing browser responsiveness and dynamic content loading without full page refreshes. The interface consists of four main pages:

\begin{itemize}
\item \texttt{\textbf{Login Page:}} Users log in with a numeric user ID. The ID is stored in local storage and validated to prevent spam at the frontend level.
\item \texttt{\textbf{Portfolio Page:}} Displays the user's portfolio as cards arranged in rows and columns, ordered from most recent to oldest.
\item \texttt{\textbf{Recommendation Page:}} Shows personalized stock recommendations as a list of buttons, allowing users to add a stock to their portfolio.
\item \texttt{\textbf{Stock Page:}} Displays detailed stock information including EPS, MACD-K, MACD-S, price, and stock name, alongside charts and side-panel recommendations. An "Add to Portfolio" button is positioned at the top of the page.
\end{itemize}
\section{Limitations}

The following limitations apply to individual system components:

\begin{itemize}
\item \texttt{\textbf{Leader ML Model:}} The leader must be fully deployed and running before any follower can start. While Google's Object Storage bandwidth is 8 Gib/s, weight writes take a maximum of 2--5 seconds.
\item \texttt{\textbf{User Limit:}} The system is limited to 1,000 users due to the \texttt{StringLookup} vocabulary size. This can be increased by adjusting a single parameter, or set to \texttt{None} for the maximum possible combinations at the current tensor length of 164.
\item \texttt{\textbf{Recommendation Request Speed:}} Each follower processes requests in 2--3 seconds due to busy-wait locks added for thread safety. Migrating to monitor-based synchronization would reduce overhead.
\item \texttt{\textbf{BigQuery Latency:}} BigQuery queries take approximately one second, limiting stock information retrieval to roughly 1.5 seconds.
\item \texttt{\textbf{Data Freshness:}} Stock data is refreshed every 150 minutes, meaning users may see data up to 45 minutes old for MACD, price, and EPS. This can be improved with a higher Alpha Vantage API plan \cite{vantage} and faster batch uploads through Kubernetes persistent storage.
\item \texttt{\textbf{Portfolio Addition Throughput:}} Followers can process up to 300 portfolio addition requests per second, but the leader requires 2--5 seconds to write each update to disk. A pre-scaled cluster or Redis cache would resolve this bottleneck in production.
\item \texttt{\textbf{Security:}} No service mesh is deployed, so inter-cluster calls lack mTLS. An attacker who penetrates the system could read internal traffic. IAM is also not configured for endpoints. Both are left as future work.
\end{itemize}

\section{Future Scope}

The model has no retraining downtime due to its hybrid architecture, except when the leader fails. Planned improvements include:
\begin{itemize}
\item \texttt{\textbf{Alpha Vantage API:}} Upgrade to 1,200 API calls per minute, reducing data latency to a few seconds.
\item \texttt{\textbf{Time-Series Storage:}} Replace BigQuery with InfluxDB for lower latency, higher throughput, and automatic data expiry.
\item \texttt{\textbf{Candidate Generation:}} Replace brute-force matrix factorization with Google's SCANN algorithm for faster retrieval, with a trade-off in accuracy.
\item \texttt{\textbf{Faster File Storage:}} Replace Google Object Storage with NFS or Redis cache for faster writes, though at higher cost and with storage size constraints that may require Redis sharding.
\item \textbf{\texttt{IAM and Service Mesh:}} Install identity and access management and a service mesh for enhanced security beyond frontend filtering.
\item \texttt{\textbf{Database Scaling:}} Expand from a single PostgreSQL instance to a leader-follower or cloning architecture to handle higher traffic.
\end{itemize}

\section{Conclusion}

In this paper, we presented a scalable online deep learning recommendation system for stock portfolio management. The system was implemented using TensorFlow Recommenders and deployed on a distributed architecture built with Kubernetes, Docker, RabbitMQ, BigQuery, and PostgreSQL. The model generates personalized stock recommendations in 2--3 seconds using five input parameters — MACD-K, MACD-S, EPS, current price, and stock name — and trains for three epochs on each new observation. The use of online learning enables the system to capture concept drift, adapting recommendations to evolving user preferences over time without requiring full retraining cycles.

The hybrid leader-follower architecture provides a practical balance between training throughput and serving availability. The leader continuously updates model weights to cloud object storage, while stateless followers serve recommendations in parallel, achieving up to 300 portfolio addition requests per second per instance. Fault tolerance is ensured through RabbitMQ message durability, Kubernetes preemption-priority scheduling, and weight persistence in Google Object Storage, allowing followers to recover seamlessly after failure.

Several limitations remain. The system suffers from a cold start problem where new users receive limited recommendations until sufficient rating data is collected. The 150-minute data refresh interval, imposed by Alpha Vantage API rate limits, introduces staleness in stock indicators. Additionally, the use of busy-wait locks for thread safety and brute-force retrieval for candidate generation constrains throughput, and the absence of a service mesh and IAM leaves internal cluster calls unsecured.

Future work will focus on upgrading to a higher-tier Alpha Vantage API plan for faster data acquisition, migrating to InfluxDB for time-series storage, adopting SCANN
for approximate nearest-neighbor retrieval, and replacing busy-wait locks with monitor-based synchronization. Deployment of a service mesh and IAM would address the current security gaps, and scaling the database from a single instance to a leader-follower replication topology would better support production-level traffic. Finally, we acknowledge the fundamental trade-off between consistency and availability inherent in this architecture — a design choice that future iterations must navigate based on application requirements.

\begin{acks}
To Suraj Shreshtha for guiding me through the process for writing the frontend.
\end{acks}

\section{Citations and Bibliographies}
\bibliographystyle{plainnat}
\bibliography{acmart}

\end{document}